\documentclass[twocolumn]{aastex63}

\usepackage{amsmath}
\usepackage{amssymb}
\usepackage{textcomp}

\newcommand{\ct}{$^{13}{\rm C}$}
\newcommand{\ctb}{$^{13}{\rm C}$~}
\newcommand{\nf}{$^{14}{\rm N}$~}

\newcommand{\ms}{$M_{\odot}$~}
\newcommand{\msb}{$M_{\odot}$~}

\shorttitle{Magnetic-buoyancy induced mixing in AGB Stars}
\shortauthors{Vescovi et al.}

\begin{document}

\title{Magnetic-buoyancy Induced Mixing in AGB Stars: Presolar SiC Grains}

\correspondingauthor{Diego Vescovi}
\email{diego.vescovi@gssi.it}

\author{Diego Vescovi}
\affil{Gran Sasso Science Institute, Viale Francesco Crispi, 7, 67100 L'Aquila, Italy}
\affiliation{INFN, Section of Perugia, Via A. Pascoli snc, 06123 Perugia, Italy}
\affiliation{INAF, Observatory of Abruzzo, Via Mentore Maggini snc, 64100 Teramo, Italy}

\author{Sergio Cristallo}
\affiliation{INAF, Observatory of Abruzzo, Via Mentore Maggini snc, 64100 Teramo, Italy}
\affiliation{INFN, Section of Perugia, Via A. Pascoli snc, 06123 Perugia, Italy}

\author{Maurizio Busso}
\affiliation{University of Perugia, Department of Physics and Geology, Via A. Pascoli snc, 06123 Perugia, Italy}
\affiliation{INFN, Section of Perugia, Via A. Pascoli snc, 06123 Perugia, Italy}

\author{Nan Liu}
\affiliation{Laboratory for Space Sciences and Physics Department, Washington University in St. Louis, St. Louis, MO 63130, USA}
\affiliation{McDonnell Center for the Space Sciences, Washington University in St. Louis, St. Louis, MO 63130, USA}

\begin{abstract}
Isotope ratios can be measured in presolar SiC grains from ancient Asymptotic Giant Branch (AGB) stars at permil-level (0.1\%) precision. Such precise grain data permit derivation of more stringent constraints and calibrations on mixing efficiency in AGB models than traditional spectroscopic observations.
In this paper we compare SiC heavy-element isotope ratios to a new series of FRUITY models that include the effects of mixing triggered by magnetic fields. 
Based on 2D and 3D simulations available in the literature, we propose a new formulation, upon which the general features of mixing induced by magnetic fields can be derived.
The efficiency of such a mixing, on the other hand, relies on physical quantities whose values are poorly constrained. We present here our calibration by comparing our model results with the heavy-element isotope data of presolar SiC grains from AGB stars. We demonstrate that the isotopic compositions of all measured elements (Ni, Sr, Zr, Mo, Ba) can be simultaneously fitted by adopting a single magnetic field configuration in our new FRUITY models.

\end{abstract}


\section{Introduction} \label{sec:intro}
Thermally-Pulsing Asymptotic Giant Branch (TP-AGB) stars are among the most efficient polluters of the interstellar medium \citep{bu99,he05,stra06,ka14}. Those objects present an onion-like structure, with a partially degenerate C-O core, surrounded by two thermonuclear shells, burning He and H alternatively, and an expanded and cool convective envelope, continuously eroded by intense mass-loss phenomena.
The products of the rich nucleosynthesis occurring in their interiors are carried to the surface via mixing episodes known as Third Dredge Up (TDU). During a TDU episode, the convective envelope penetrates through the H-shell, which is temporarily switched off due to the expansion triggered by the occurrence of a thermonuclear runaway, named Thermal Pulse (TP). 
In AGB modelling, particularly critical is the handling of the convective/radiative interface at the inner border of the convective envelope, whose numerical treatment has dramatic consequences on both the efficiency of TDU and the nucleosynthesis of heavy elements in those objects. AGB stars are the site of the main component of the \textit{slow} neutron capture process (\textit{s}-process; see e.g. \citealt{ga98}). The major neutron source in AGB stars is the $^{13}$C($\alpha$,n)$^{16}$O reaction (see e.g. \citealt{cri18}), which burns in radiative conditions during the interpulse phase between two TPs \citep{stra95}. A $^{13}$C-enriched layer is needed to reproduce the observed \textit{s}-process distributions: the so-called \ctb pocket \citep{buss01}. In order to obtain the \ctb pocket, a partial mixing of hydrogen from the envelope to the underlying radiative He-intershell is needed during a TDU episode. Various mechanisms for causing this partial mixing have been proposed in stellar evolutionary codes: diffusive overshoot \citep{herw97}, rotation \citep{he03,si04}, gravity waves \citep{deni03}, opacity-induced overshoot \citep{cris09}, and a combination of overshoot and gravity waves \citep{umbi}.
None of these treatments, however, have been able to simultaneously reproduce all the \textit{s}-process isotopic anomalies measured in presolar SiC grains in detail (see \citealt{zinner} for a review). 
Presolar SiC grains have been identified in pristine extraterrestrial materials that formed shortly after the solar system birth (about 4.57 Gyr ago), and have remained intact and almost unaltered until the present day. Extensive analyses of presolar SiC grains for their multi-element isotopic compositions show that the majority ($\simeq$ 90\%), the so-called mainstream (MS) grains, came from low-mass C-rich AGB stars and exhibit \textit{s}-process isotopic signatures.
Recently, the idea that the formation of the \ctb pocket can be induced by magnetic buoyancy has been proposed by \citealt{trip16} (based on the formalism presented by \citealt{nucc14}). 
Such a treatment has been proven to be effective in reproducing many of the features characterizing \textit{s}-process distributions \citep[see][]{trip16,palm18,vesc18}. The inclusion of this process in AGB stellar models, however,  is currently confined to post-process techniques. 
In this Letter, we present our implementation of mixing triggered by magnetic buoyancy in the FUNS stellar evolutionary code with fully coupled nucleosynthesis \citep{stra06,cris11,pier13}.

\section{Updated FRUITY Models} \label{sec:impro}
\begin{figure*}[!t]
\centering
\includegraphics[width=\textwidth]{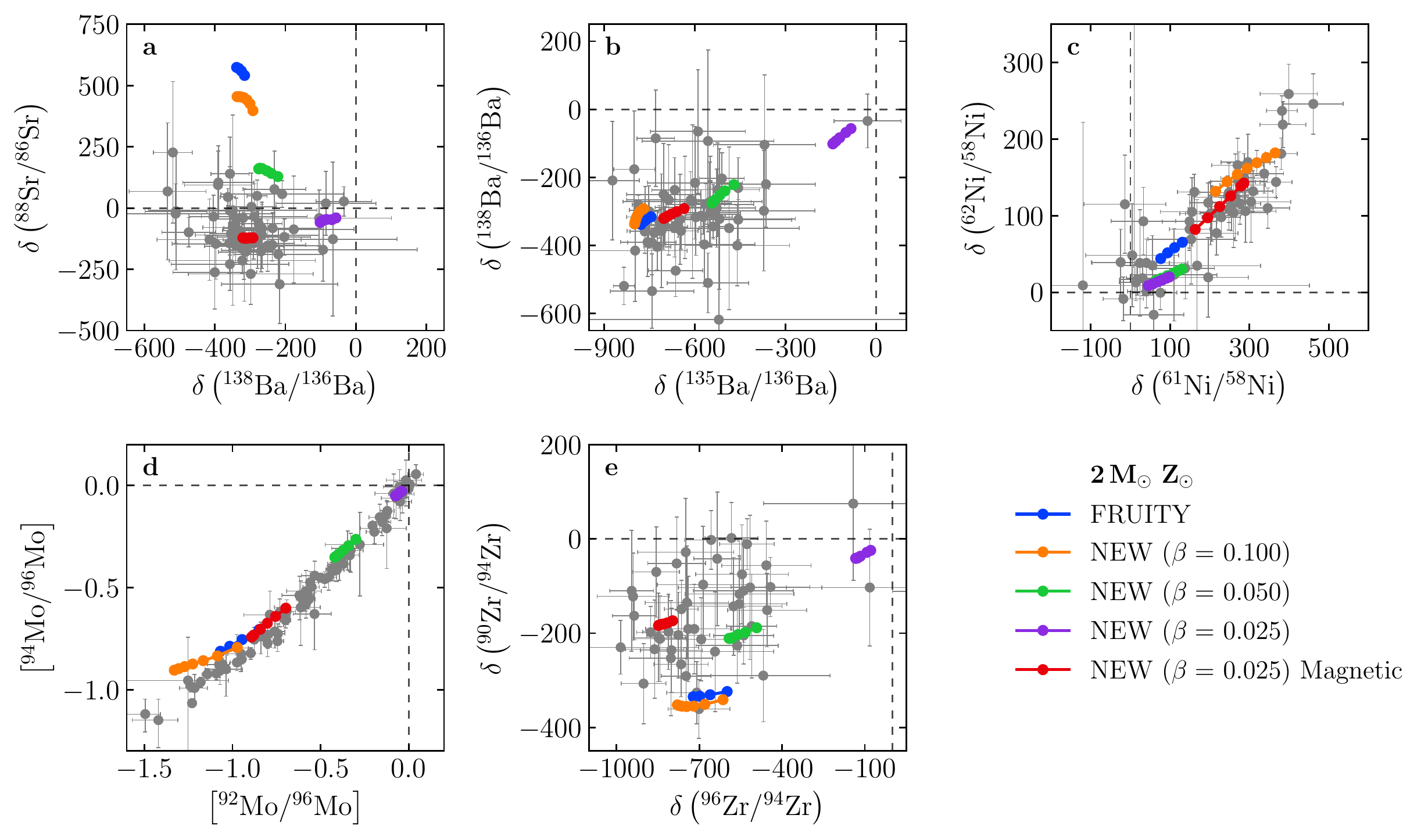}
\caption{Comparison between presolar grain data (see text for details) and theoretical stellar predictions calculated under different physical prescriptions. Plotted are 2$\sigma$ errors.}
\label{fig:grains}
\end{figure*} 
Besides the inclusion of mixing triggered by magnetic fields, we made several updates to the FUNS code. Those include the initial solar-scaled composition \citep{lodd20}, the mass-loss law \citep{abia2020}, the Equation Of State (EOS) and many nuclear reaction rates  (for these last items we refer to Vescovi \& Cristallo, {\it in preparation}). 


\subsection{Convective Overshooting} \label{sec:coos}

In FRUITY models previously reported (\citealt{cris11,pier13,cris15} available on-line in the FRUITY repository\footnote{\url{http://fruity.oa-teramo.inaf.it/}}),
the interface at the inner border of the convective envelope is handled by introducing an exponentially decaying profile of the convective velocities. 
The velocity of the descending material (ideally braked by viscous friction) appears as:
\begin{equation} \label{eq:vconv}
v = v_{\rm cb} \, e^{ -k \delta r } \, ,
\end{equation}
where $v_{\rm cb}$ is the velocity at the convective border and $\delta r$ is the corresponding distance.
It is common to assume that the convective zone extends over a fraction $\beta H_{\rm p}$ beyond the Schwarzschild's limit, where $H_{\rm p}$ is the pressure scale height at the convective boundary \citep[see e.g.][]{frey96}: for this reason we define $k = (\beta H_{\rm p})^{-1}$. 
The free parameter $\beta$ regulates the amount of protons mixed beyond the bare convective border, and also affects the TDU efficiency. 
The introduction of Eq.~\ref{eq:vconv} has an important by-product, i.e. the formation of a self-consistent $^{13}$C pocket, whose size decreases with the shrinking of the He-intershell. 
\citet{cris09} tuned $\beta$ to maximize the production of \textit{s}-process elements ($\beta$ = 0.1). AGB models computed with this value have proven to be effective in roughly reproducing the bulk of the luminosity function of Galactic C-stars \citep{guan13} and the solar distribution of \textit{s}-only isotopes \citep{pra2020}. For this reason we define the FRUITY models computed with $\beta$ = 0.1 
as our reference scenario for what concerns the TDU efficiency (blue symbols in Fig.~\ref{fig:grains}).

In order to evaluate the effects induced by different physical recipes in calculating AGB models, we ran a series of \textit{s}-process AGB models with an initial mass M = 2 \ms and Z = $1.67\times 10^{-2} (\equiv$ Z$_\odot$). We compared the model results to the isotopic ratios of $s$-elements in presolar SiC grains, which offer precise constraints on the \ctb pocket (see Fig.~\ref{fig:grains}). We included MS grain data for Ni \citep{trap18}, Sr \citep{liu15,step18}, Zr \citep{nico97,barz07}, Mo \citep{liu17,step19} and Ba \citep{liu14a,liu15,step18}. 
We also included the Mo isotopic compositions of presolar SiC grains of types Y and Z from \cite{liu19}, because their Mo isotopic compositions have been demonstrated to be indistinguishable from those of MS grains. 
Although observations show that C-rich dust can sometimes form in O-rich circumstellar  envelopes \citep[see e.g.][]{mill16}, we conservatively plot the model data only for the C-rich phase, during which SiC grains most likely form \citep[see also][]{lodd99}.
The presolar SiC data for all elements but Mo are reported in the typical $\delta$-notation, i.e. the deviation in parts per thousand of the isotopic ratio measured in a grain relative to the terrestrial ratio. The Mo isotope data are presented in the usual spectroscopic notation\footnote{[A/B]=log(N(A)/N(B))$_*$-log(N(A)/N(B))$_\odot$}. In Fig.~\ref{fig:grains}a we focus on $^{88}$Sr and $^{138}$Ba, both of which have magic numbers of neutrons (N = 50 and N = 82, respectively). As a consequence, they act as bottlenecks of the \textit{s}-process and are the most representative isotopes for light and heavy \textit{s}-elements (\textit{ls} and \textit{hs}, respectively). 
In addition, correlations among them were shown to depend strongly on the extension of the \ctb reservoir and on the profile of the \ctb abundance within the pocket \citep[see e.g.][]{liu15}. 
From Fig.~\ref{fig:grains} it clearly emerges that the reference 
FRUITY model has serious problems in reproducing the presolar grain isotopic ratios. 
In particular, this model predicts too high $\delta(^{88}$Sr/$^{86}$Sr) and relatively low $\delta(^{90}$Zr/$^{94}$Zr) values, resulting in poor fits to the grains in Fig.~\ref{fig:grains}a and \ref{fig:grains}e. In fact, both isotope ratios were shown to be sensitive tracers of the \ctb pocket structure \citep{liu14b,liu15}.
Regarding Ni isotopes, the predicted \textit{s}-process enrichments in $^{61}$Ni and $^{62}$Ni in the envelope fail to explain the grains with the largest $\delta$ values.
The poor match to the grain data is barely improved by the inclusion of the new inputs (initial composition, mass-loss, EOS and nuclear rates; label ``NEW ($\beta$ = 0.100)''), apart from a net improvement for the most anomalous grains in Fig.~\ref{fig:grains}c. 
An inspection of Fig.~\ref{fig:grains}, however, is not sufficient. In addition to relative isotopic ratios, the absolute amount of freshly synthesized elements also has to be checked.
 
\begin{deluxetable*}{lccccc}[t!]
\label{tab:models}
\tablecaption{Final enrichments of \textit{s}-process elements and \textit{s}-process indexes. See text for details.}
\tablehead{
\colhead{} & \multicolumn5c{Models} \\
\cline{2-6}
\colhead{} & \colhead{FRUITY} & \colhead{NEW} & \colhead{NEW} & \colhead{NEW} & \colhead{NEW Magnetic} 
}
\startdata
$Z_\odot$ & 0.0138 & 0.0167 & 0.0167 & 0.0167 & 0.0167 \\
$\beta$ & 0.100 & 0.100 & 0.050 & 0.025 & 0.025 \\
\hline
\hline
$\Delta{\rm M_{TDU}}$ & 2.91$\times 10^{-2}$ & 6.20$\times 10^{-2}$ & 4.15$\times 10^{-2}$ & 3.28$\times 10^{-2}$ & 3.27$\times 10^{-2}$  \\
\hline
\text{Yield(C)} & 5.24 $\times 10^{-3}$ & 9.75 $\times 10^{-3}$ & 6.51 $\times 10^{-3}$ & 5.23 $\times 10^{-3}$ & 5.17 $\times 10^{-3}$ \\
\text{Yield(Ni)} & 1.61 $\times 10^{-7}$ & 1.37 $\times 10^{-6}$ & -1.61 $\times 10^{-7}$ & -1.63 $\times 10^{-7}$ & 1.36 $\times 10^{-6}$ \\
\text{Yield(Sr)} & 5.81 $\times 10^{-7}$ & 9.61 $\times 10^{-7}$ & 8.12 $\times 10^{-8}$ & 1.07 $\times 10^{-8}$ & 5.85 $\times 10^{-7}$ \\
\text{Yield(Zr)} & 2.72 $\times 10^{-7}$ & 4.66 $\times 10^{-7}$ & 3.59 $\times 10^{-8}$ & 3.66 $\times 10^{-9}$ & 1.76 $\times 10^{-7}$ \\
\text{Yield(Mo)} & 4.21$\times 10^{-8}$ & 7.60$\times 10^{-8}$ & 5.73$\times 10^{-9}$ & 5.77$\times 10^{-10}$ & 2.36$\times 10^{-8}$ \\
\text{Yield(Ba)} & 1.72$\times 10^{-7}$ & 3.15$\times 10^{-7}$ & 2.02$\times 10^{-8}$ & 1.43$\times 10^{-9}$ & 6.44$\times 10^{-8}$ \\
\hline
\text{[ls/Fe]}\tablenotemark{a} & 1.02 & 1.22 & 0.37 & 0.06 & 0.95 \\
\text{[hs/Fe]}\tablenotemark{b} & 0.96 & 1.17 & 0.27 & 0.02 & 0.60 \\
\text{[hs/ls]}\tablenotemark{c} & -0.06 & -0.05 & -0.10 & -0.04 & -0.35 \\
\hline
\enddata
\tablenotetext{a}{[ls/Fe]=([Sr/Fe]+[Y/Fe]+[Zr/Fe])/3}
\tablenotetext{b}{[hs/Fe]=([Ba/Fe]+[La/Fe]+[Ce/Fe]+)/3}
\tablenotetext{c}{[hs/ls]=[hs/Fe]-[ls/Fe]}
\end{deluxetable*}
In Table~\ref{tab:models} we report, for the computed models, the following quantities: amount of dredged-up material, net yields\footnote{Net yields are defined as $\int_{0}^{t_{end}}$[(X(El)-X$_0$(El))$\times \dot M$]dt, where t$_{end}$ is the stellar lifetime, $\dot M$ is the mass-loss rate, while X(El) and X$_0$(El) stand for the current and the initial mass fraction of the element, respectively.}
of some key elements and \textit{s}-process indexes.
As we already stressed before, 
FRUITY models have been demonstrated to be able to grossly reproduce Galactic chemical features.
With respect to the 
FRUITY model, the ``NEW ($\beta = 0.100$)'' model carries to the surface too much material (see, e.g., the carbon net yield), thus pointing to the need of reducing the mixing efficiency. 
The test with an intermediate $\beta$ value of 0.05 improves the situation for both the grains and net element production, but the achieved improvement is still insufficient (due to the fact that the extra-mixed region is too $^{14}$N-rich).
Therefore, we ran an additional model with $\beta=0.025$. 
The ``NEW ($\beta=0.025$)'' model shows an amount of dredged-up material similar to the reference 
FRUITY model\footnote{It is worth stressing that we could perform a finer calibration, but we believe it is premature at the moment. We will investigate this matter as soon as detailed 3D hydrodynamic simulations of an AGB penetrating envelope become available.}, but the production of heavy elements is completely suppressed in this case. This is confirmed by both the close-to-solar values in all the presolar grain isotopic ratios (violet symbols in Fig.~\ref{fig:grains}) and the extremely reduced net yields. Thus, we conclude that the new FRUITY models, re-calibrated after the inclusion of updated physical inputs, cannot  reproduce the amounts of heavy elements required by observations. This calls for an additional mechanism for the production of heavy elements.

\section{Mixing triggered by Magnetic Buoyancy in AGB stars} \label{sec:imple}

\begin{figure}[t!]
\centering
\includegraphics[width=\columnwidth]{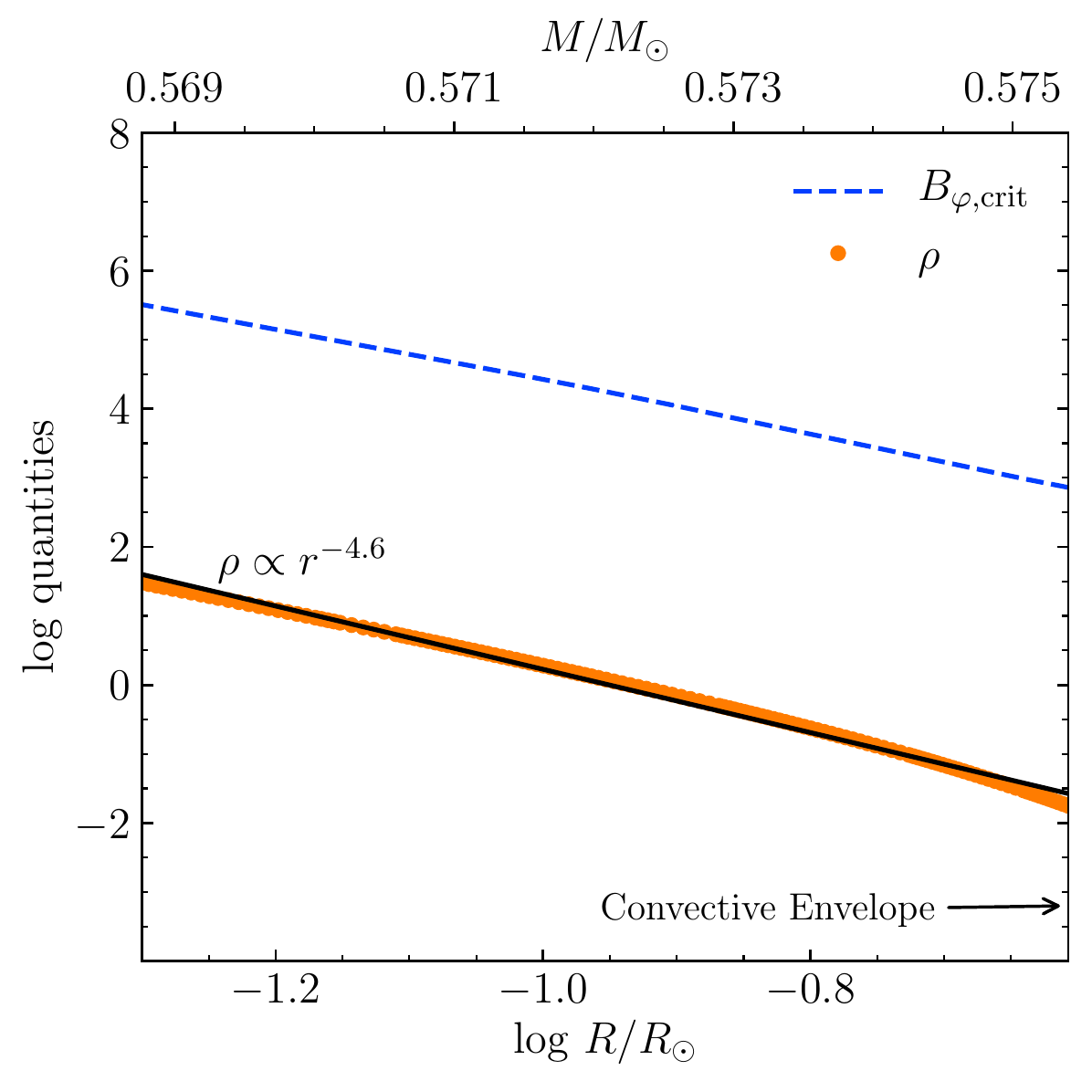}
\caption{Critical toroidal field $ B_{\varphi}$ for triggering the buoyancy instability (blue dashed curve) and density $\rho$ (orange dots) profiles of the He-intershell up to the inner border of the convective envelope (right edge of the plot), at the epoch of its maximum penetration during the 3rd TDU of a star with 2 $M_\odot$ and solar metallicity. Note that the best fit closely follows $\rho \propto r^{k}$, with $k \simeq -4.6$, which is considerably lower than $-1$, as required for the validity of the model by \cite{nucc14}.}
\label{fig:field}
\end{figure} 

As in \citet{nucc14}, we assume that a toroidal magnetic field is present in the radiative He-intershell region at the beginning of the TDU. We will demonstrate the validity of this assumption in a dedicated forthcoming paper (Vescovi \& Cristallo, {\it in preparation}).
Here we just briefly note that differential rotation may create a strong enough toroidal field (B $\sim 10^5$ G) by stretching a small preexisting poloidal field around the rotation axis (see e.g. \citealt{deni09}). The poloidal field does not need to be preserved from previous phases, since its required small strength (10 $\div$ 100 G) can be linked to a local process (such as a convective episode). The energy budget to develop and maintain such a toroidal magnetic field is provided by rotation. Preliminary tests computed by switching on rotation in our new models (see \citealt{pier13} for details) confirm the above-reported statements, even when hypothesizing a large decrease of the core rotation velocity in pre-AGB evolutionary phases (see, e.g., \citealt{denha}). In such a situation, mixing triggered by secular rotation instabilities is negligible. 

\cite{nucc14} pointed out that a magnetized stellar plasma in quasi-ideal MHD regime, with a density distribution closely following a power law as a function of the radius ($\rho \propto r^{k}$, with $k<-1$), reaches a dynamic equilibrium and is in radial expansion. The result above is analytically exact and remains so (for the simple but rather typical symmetry adopted by the authors) when the magnetic field \textit{B} varies in time, as in the case of a toroidal/azimuthal magnetic field amplified by winding-up. Here we assume that magnetic buoyancy is the instability which triggers the plasma expansion. Note that the region below the convective envelope during a TDU fullfills the conditions required by \citealt{nucc14}, with $\rho \propto r^{-4.6}$ (solid line in Fig.~\ref{fig:field}). Moreover, the occurrence of buoyancy instability requires quite strong fields (dashed line in Fig.~\ref{fig:field}) and, in such conditions, the magnetic field tends to concentrate in bundles of field lines that are wrapped in a field-free plasma, usually referred to as flux tubes. 
As a consequence of the magnetic extra-pressure, these tubes are buoyant \citep[see e.g.][]{park55}.
Due to the effect of the magnetic buoyancy, a matter flow is pushed from the He-intershell to the envelope. This, in turn, induces a downflow flux, in order to guarantee mass conservation.

A brief outline of the general downflow velocity profile we adopted is presented in Appendix~\ref{app:downflow}.
\begin{figure}[t!]
\centering
\includegraphics[width=\columnwidth]{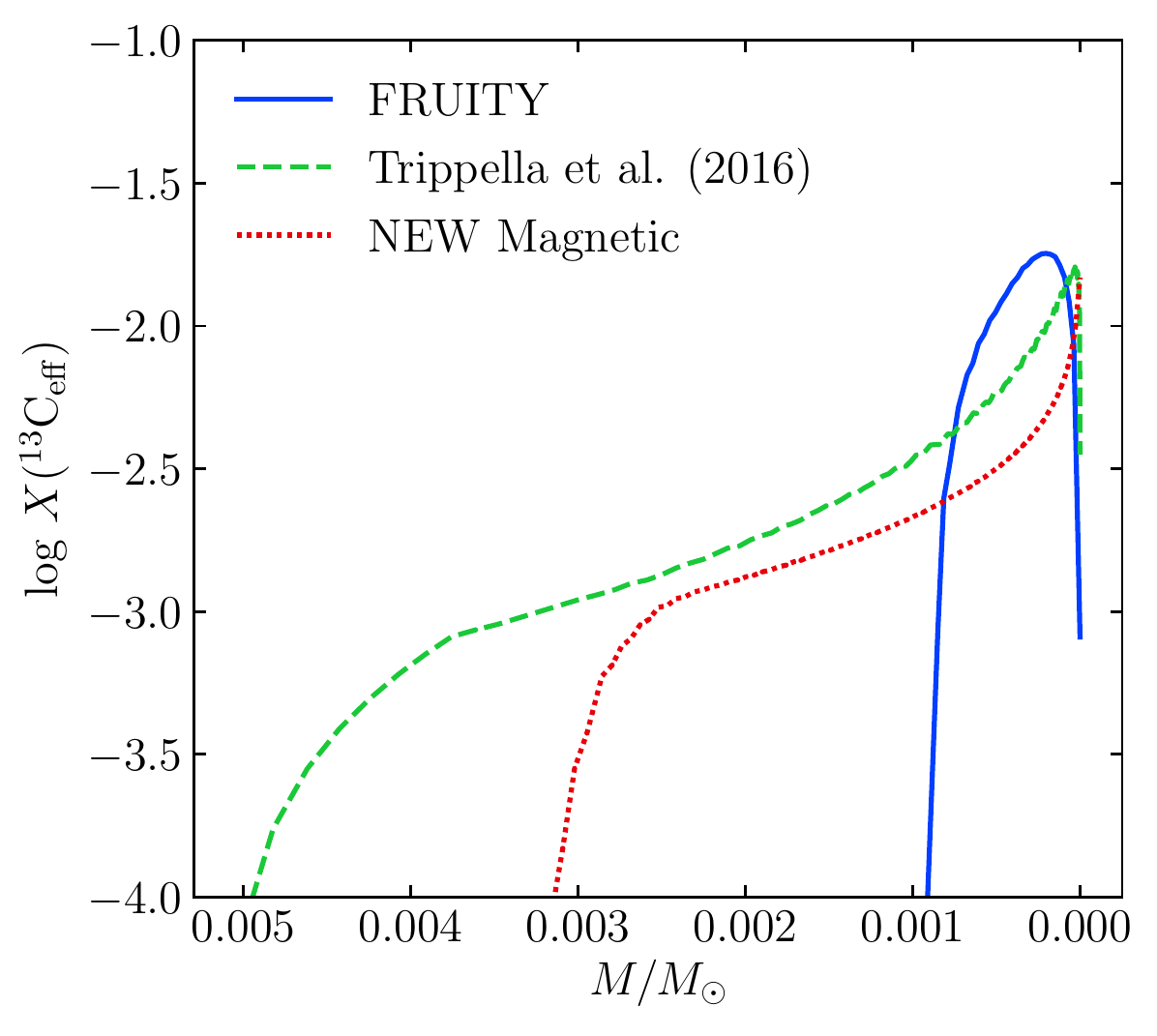}
\caption{\textit{Effective} \ctb in the \ctb pocket region for models with different physical prescriptions. See text for details.}
\label{fig:pockets}
\end{figure} 
Supposing that magnetic flux tubes, generated in the the He-intershell at a distance $r_{\rm p}$ from the stellar center, start to rise with an initial velocity $v_{\rm p}$, then the induced downflow velocity can be expressed (see Eq.~\ref{eq:vfinal}) as
\begin{equation}\label{eq:vfinal1}
v_{d}(r) = u_{\rm p}  \left( \frac{r_{\rm p}}{r} \right)^{k+2} \, ,
\end{equation}
where $u_{\rm p} = f \cdot v_{\rm p}$ acts as an \textit{effective} buoyant velocity.
In fact, in radiative zones of evolved stars, the fraction of mass \textit{f} locked in magnetic flux tubes must be small, i.e. $f\simeq 10^{-5}$ \citep[see][]{buss07,trip16}. This fact implies that the actual buoyant velocity of the flux tubes is orders of magnitude larger than the corresponding $u_{\rm p}$. 
The downflow velocity relies on two parameters: the radial position $r_{\rm p}$ of the layer p from which buoyancy (on average) starts and the \textit{effective} buoyant velocity $u_{\rm p}$. This is a direct consequence of the solutions derived by \cite{nucc14} for the radial velocity of magnetized structures and also the toroidal component of the magnetic field, as for both these functions  we need to fix boundary conditions.
\begin{figure*}[t!] 
\centering
\includegraphics[width=\textwidth]{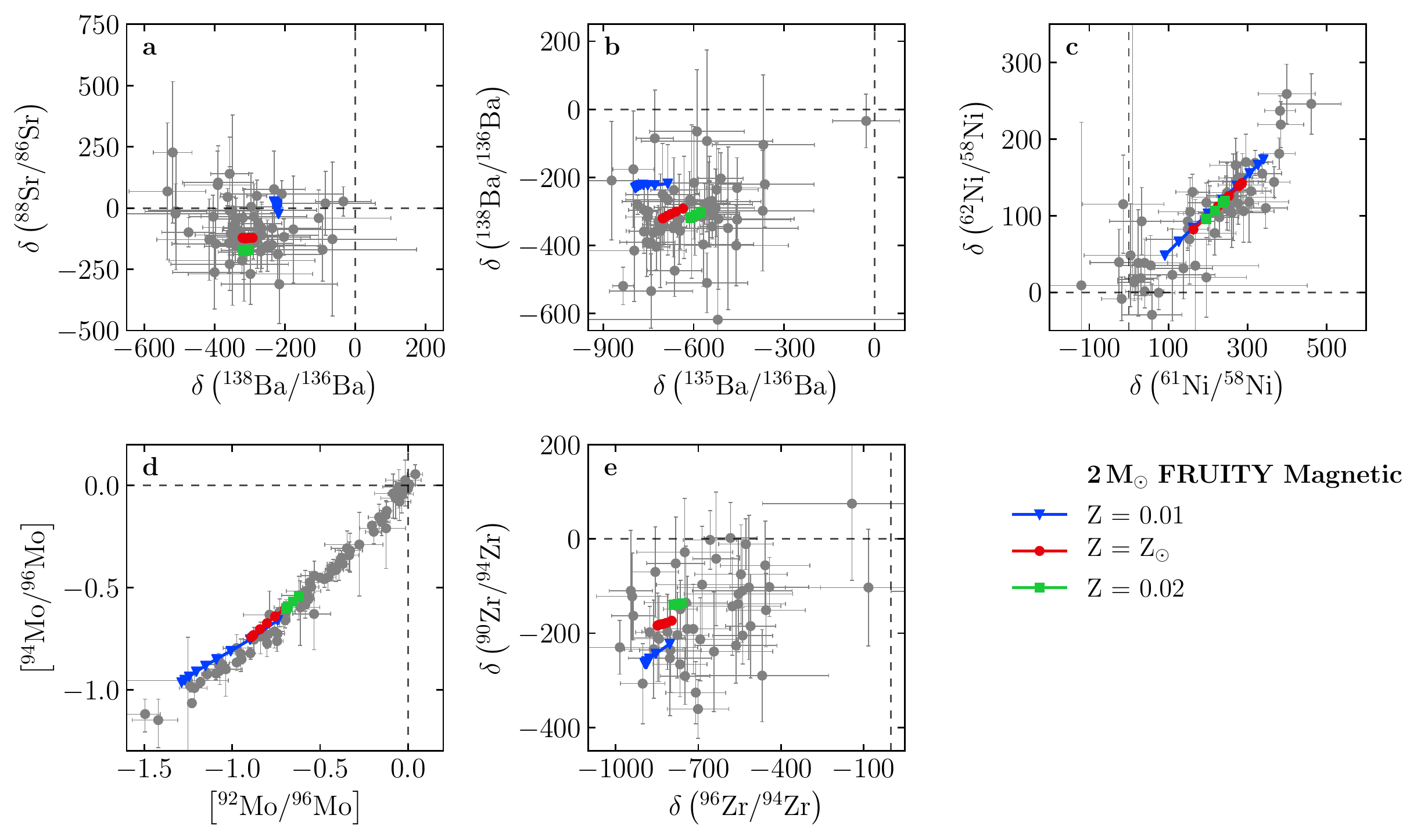}
\caption{
Same as Fig.~\ref{fig:grains}, but with new magnetic 2 \msb models at different metallicities.
See text for details.}
\label{fig:grains2}
\end{figure*} 

The identification of the critical field necessary for the occurrence of instabilities by magnetic buoyancy allows us to identify the corresponding radial position $r_{\rm p}$ from which magnetic structures arise.
An azimuthal field $B_{\varphi}$ is subject to magnetic buoyancy instabilities \citep{ache78,spru99,deni09} if:
\begin{equation}
B_{\varphi} \gtrsim \left( 4 \pi \rho r N^2 H_{\rm p} \dfrac{\eta}{K} \right)^{1/2} \, ,
\end{equation}
provided that the field gradient is smooth, i.e. $\partial \, {\rm ln} \, B_{\varphi} / \partial \, {\rm ln} \, r \sim O(1)$. Here $N$ is the adiabatic Brunt-V\"ais\"al\"a frequency, $\eta$ the magnetic diffusivity and $K$ the thermal diffusivity.

Fig.~\ref{fig:field} shows the profile of the critical $B_{\varphi}$ necessary for the onset of  magnetic buoyancy instabilities, in the radiative zone below the convective envelope, at the moment of the maximum penetration of the H-rich envelope during a TDU.
$B_{\varphi}$ varies from $\sim 10^4$ G to a few $10^5$ G, in the region of interest for the formation of the \ctb pocket. 
Different values for $B_{\varphi}$ correspond to different values of the free parameter $r_{\rm p}$, which determines the extension of the mixed zone and, in turn, of the \ctb pocket. Assuming that a fraction of the magnetic energy is converted to the kinetic energy of the magnetic flux tube, we expect that the (\textit{effective}) rising velocity of the flux tubes is proportional to the strength of the magnetic field ($u_{\rm p} \propto v_{\rm p} \propto B_{\varphi}$). To calibrate $u_{\rm p}$ and $B_{\varphi}$, we ran various tests with different parameter values ($u_{\rm p}$ = 1, 3, 5, 8, 12 $\times$ $10^{-5}$ cm s$^{-1}$ and $r_{\rm p}$ corresponding to $B_{\varphi}$ = 2, 5, 10, 15 $\times$ 10$^{4}$ G).
From eq.~\ref{eq:vfinal1} it is straightforward to notice that the velocity of the downward material is proportional to $v_{\rm p} r_{\rm p}^{k+2}$ (with \textit{k} typically \textless $-4$, during a TDU). Thus, the greater the initial velocity of flux tubes is and the deeper the buoyancy starts, the greater the velocity of the material down-flow is.
Therefore, larger values of $B_{\varphi}$ correspond both to larger \ctb pockets and to larger mass fractions $X$(\ct). 
The case that provides the best fit to the presolar SiC grain isotopic ratios was obtained with $B_{\varphi} = 5 \times 10^{4}$ G and $u_{\rm p} = 5 \times 10^{-5}$ cm s$^{-1}$ (red symbols in Fig.~\ref{fig:grains}).
This $u_{\rm p}$ value corresponds to a starting buoyant velocity of $v_{\rm p} = u_{\rm p} / f \simeq$ 5 cm s$^{-1}$, which increases to $\simeq$ 5 m s$^{-1}$ at the convective boundary, thus ensuring that magnetic advection acts on timescales much smaller than any dissipative processes \citep{nucc14}.

Fig.~\ref{fig:pockets} shows the amount of \textit{effective} \ctb (i.e. the difference between the number fractions of \ctb and \nf in the pocket)
obtained by including our new magnetic mixing process after the 3$^{rd}$ TDU in a 2 $M_\odot$ model with Z = Z$_\odot$ (red dotted curve). For comparison, we also show typical \ctb pockets obtained in 
FRUITY models (blue solid curve)
and the ``magnetic'' pocket by \citet{trip16} (green dashed curve). With respect to the 
FRUITY pocket, our new ``Magnetic'' pocket shows a lower $^{13}$C concentration, with a more extended tail. Our formulation shares the physical principles of \citet{trip16}. 
Adopting the same post-process code by those authors, we have performed a large series of computations to investigate the variability of the magnetic-buoyancy induced pocket. The results will be presented elsewhere (Busso et al., \textit{in preparation}).
We briefly note that in advanced cycles the \ctb pockets shrink somewhat, in almost perfect agreement with the trend observed in the models presented here.
%
\section{Presolar SiC grains} \label{secsic}
In order to test if the new magnetic FRUITY AGB models are able to cover the range defined by the presolar SiC grains, we computed two additional 2 \msb AGB evolutionary sequences with Z = $1\times 10^{-2}$ and Z = $2\times 10^{-2}$.
In Fig.~\ref{fig:grains2}, those models are compared to the same set of data as in Fig.~\ref{fig:grains}.
The inclusion of the ``Magnetic'' \ctb pocket in our new FRUITY models significantly improves the fits to the majority of the grain data in all the panels in Fig.~\ref{fig:grains2}. This is especially true in Fig.~\ref{fig:grains2}a: all the model predictions overlap well with the grain region, in contrast to the poor matches given by the FRUITY models adopting other \ctb pockets shown in Fig.~\ref{fig:grains}a.
The Z = 0.01 model predicts larger \textit{s}-process enrichments in the envelope because of the increased neutron-to-seed ratio, thus providing better matches to the grains with extreme Ni and Mo isotopic anomalies.
The \textit{s}-process features detected in presolar grains are, therefore, well reproduced by including our treatment of magnetic-buoyancy induced mixing in our AGB models.
It was shown that grain data for Ni, Zr, and Mo suffer from solar and/or terrestrial contamination. This drives their composition toward more normal values and likely results in the large spreads observed in panels c-d-e of Fig.~\ref{fig:grains2}  \citep{trap18,liu18,liu19}. Given this caveat, we focus on matching the most anomalous grains in these three cases: the less anomalous grains can be explained by mixing our predicted s-process components with solar and/or terrestrial materials. 
Alternatively, the less anomalous grains could probably be explained by the models shown in Fig.\ref{fig:grains2}, if a higher carbon content in the He-intershell is adopted so that the stellar envelope becomes C-rich earlier (see e.g. \citealt{batt19}).
In our analysis, we implicitly assumed that the initial mass (2 $M_\odot$) and metallicities of the computed models are representative of the population of grain parent stars: this choice is commonly adopted in the literature (see, e.g., \citealt{lewis13} and \citealt{liu18}). However, it has been recently proposed that more massive AGB stars (M $\simeq 4~M_\odot$) with super-solar metallicities (Z = 2 $\times$ Z$_\odot$) are the parent stars of presolar SiC grains \citep{lug18}. We will investigate this subject in a dedicated paper
(Cristallo et al., {\it in preparation}). \\
Finally, an important caveat needs to be noted. Theoretically, the physical requirements given by \citet{nucc14} to ensure quasi-ideal MHD conditions still hold slightly deeper inside the star, with respect to the adopted configuration, down to layers with a larger critical magnetic field $B_{\varphi}$. Currently, we do not have the means to perform the absolute calibration of the free parameters in our treatment of MHD-induced mixing, and their values were calibrated using the presolar grain data.  However, we anticipate that the calibrated values from this study also allow us to obtain an overall reasonable fit to the surface distributions determined in other \textit{s}-process enriched objects, including intrinsic C-stars, Ba-stars, CH stars and CEMP-s stars.
As a matter of fact, we have hints that the observed  \textit{s}-process spread at a fixed metallicity is connected to the initial mass and/or rotational velocity of the star. 
According to model simulations, mixing triggered by rotation-induced instabilities is (almost) inhibited in AGB stars that are slowed down to match asterosesmic asteroseismic measurements of core H- and He-burning stars \citep{pier13,denh19b}.
Notwithstanding, the residual angular velocity profile keeps memory of the assigned initial parameters and of the following pre-AGB evolution.
All these features will be addressed in a forthcoming paper (Vescovi \& Cristallo, {\it in preparation}).

\section{Conclusions}\label{conclu}
In this Letter we presented the first numerical simulations of the formation of a magnetically-induced \ctb pocket in a stellar evolutionary code with fully coupled nucleosynthesis. We propose that magnetic fields of the order of $10^5$ G can induce the formation and buoyant rise of magnetic flux tubes in the He-intershell of AGB stars. Such tubes are fast enough to guarantee, by mass conservation, the downward penetration of a sufficient protons to form a sizable \ctb pocket. 
With a proper choice of the field strength and initial buoyant velocity, our new magnetic FRUITY models provide  a consistent explanation to the majority of the heavy-element isotope data detected in presolar SiC grains from AGB stars.

\acknowledgments

We thank a skilled referee, who largely improved the quality of this paper. D.V. and S.C. deeply thank Luciano Piersanti for keen suggestions on the draft and for a fruitful long-lasting collaboration. 
N.L. acknowledges financial support from NASA (80NSSC20K0387 to N.L.).

\software{FUNS \citep{stra06,cris11,pier13}}

\appendix

\section{Downflow due to magnetic buoyancy}\label{app:downflow}
To derive the downflow velocity profile, we hypothesize that a magnetic flux torus, of radius $a(r_{\rm p})$, which formed in the He-intershell region due to the kink-mode buoyancy instability, starts to buoy at a distance $r_{\rm p}$ from the stellar center and reaches the H-rich material of the envelope at $r_{\rm h}$. Its volume is $V(r_{\rm p})= 2\pi^2 a^2(r_{\rm p}) r_{\rm p}$.
For the mass conservation within the flux tube (isolated matter) $\rho(r_{\rm p})V(r_{\rm p})=\rho(r_{\rm h})V(r_{\rm h})$, one has
\begin{equation} \label{A1}
\rho(r_h)= \rho(r_{\rm p})\frac{V(r_{\rm p})}{V(r_{\rm h})} = \rho(r_{\rm p})\frac{a^2(r_{\rm p})}{a^2(r_{\rm h})}\frac{r_{\rm p}}{r_{\rm h}} \, .
\end{equation}
From the magnetic flux conservation one derives
\begin{equation} \label{A2}
\frac{B_{\varphi} (r_{\rm h})}{B_{\varphi}(r_{\rm p})} = \frac{a^2(r_{\rm p})}{a^2(r_{\rm h})} \, .
\end{equation}
If the density of radiative layers below the convective envelope of an evolved star drops with the radius as a power law (i.e. $\rho (r) \propto r^k$, with an exponent k that is negative and has a modulus larger than unity), then the toroidal magnetic field can be expressed as $B_{\varphi}(r)=B_{\varphi}(r_{\rm p})(r/r_{\rm p})^{k+1}$ (see Appendix in \citealt{nucc14}). From Eq.~\ref{A2}, it follows that 
\begin{equation}
\frac{B_{\varphi}(r_{\rm h})}{B_{\varphi}(r_{\rm p})} = \frac{a^2(r_{\rm p})}{a^2(r_{\rm h})} = \left( \frac{r_{\rm h}}{r_{\rm p}} \right) ^{k+1} \, .
\end{equation}
Thus Eq.~\ref{A1} becomes
\begin{equation} \label{A4}
\rho(r_{\rm h})=\rho(r_{\rm p}) \left( \frac{r_{\rm h}}{r_{\rm p}} \right) ^{k} \, .
\end{equation}
If we consider that the magnetized regions will occupy a fraction $f(r)$ of the total mass of a stellar layer of radius $r$ \citep{trip16},
the rate of the total rising mass is $\dot{M}(r_{\rm p})=4 \pi r_{\rm p}^2 \rho(r_{\rm p}) v(r_{\rm p}) f(r_{\rm p})$.
Let's assume that the velocity of the rising flux tubes varies as \citep{nucc14}
\begin{equation}\label{A5}
v(r) = v(r_{\rm p}) \left( \frac{r_{\rm p}}{r} \right) ^{k+1} \, .
\end{equation} 
Then, the mass flow at $r_{\rm h}$ would be 
\begin{equation}
\dot{M}(r_{\rm h}) = 4 \pi r_{\rm h}^2 \rho(r_{\rm h}) v(r_{\rm h}) f(r_{\rm h})
= 4 \pi r_{\rm h}^2 \rho(r_{\rm p}) \left( \frac{r_{\rm h}}{r_{\rm p}} \right)^{k} v(r_{\rm p}) \left( \frac{r_{\rm p}}{r_{\rm h}} \right) ^{k+1} f(r_{\rm h}) \, .
\end{equation}
Being the rising mass conserved in its upward motion ($\dot{M}$ = constant), this implies that $r \cdot f(r)$ = constant, and therefore
\begin{equation}\label{A6}
    f(r_{\rm h}) = f(r_{\rm p}) r_{\rm p} / r_{\rm h} \, .
\end{equation}
Maintaining mass conservation across the envelope requires that 
$ v(r_{\rm h})f(r_{\rm h})=v_{d}$, where $v(r_{\rm h})$ is the velocity of the buoyant flux tubes at $r_{\rm h}$, and $v_{d}$ is the initial velocity of envelope material injected into the He-rich layers.
From equations~\ref{A4},\ref{A5},\ref{A6}, one obtains 
\begin{equation}
v_{d}= v(r_{\rm p}) f(r_{\rm p}) \left( \frac{r_{\rm p}}{r_{\rm h}} \right)^{k+2} \, .
\end{equation}

Since the density of radiative layers below the convective envelope has a density distribution of the form $\rho (r) \propto  r^k$, 
considering mass conservation, it is possible to  write the velocity dependency on the radius as $v_{d}(r)= v_{d} \left(r_{\rm h}/r \right) ^{k+2}$. We finally derive
\begin{equation} \label{eq:vfinal}
v_{d}(r) = u_{\rm p}  \left( \frac{r_{\rm p}}{r} \right)^{k+2} \, ,
\end{equation}
where we set $u_{\rm p}$ $\equiv$ $v(r_{\rm p})f(r_{\rm p})$.


\bibliographystyle{aasjournal} 
\bibliography{biblio}

\end{document}